# Guildenstern and Rosencrantz in Quantumland
# A Reply to Adrian Kent

Vincent Duhamel, Paul Raymond-Robichaud

**Abstract :**

This paper is an answer to the first part of Adrian Kent's *One World versus Many : the Inadequacy of Everettian Accounts of Evolution, Probability, and Scientific Confirmation*. We take issue with Kent's arguments against many-world interpretations of quantum mechanics. We argue that his reasons for preferring single-world interpretations are logically flawed and that his proposed single-world alternative to probability theory suffers from conceptual problems. We use a few thought-experiments which show that the problems he raises for probabilities in multiverses also apply in a single universe.

**Prologue**

The publication of Everett's doctoral thesis has led to a variety of debates, both in physics and the philosophy of physics, debates which focus on whether or not one can actually make sense of Everettian interpretations of quantum mechanics. Among other problems, single-universe proponents challenge many-worlders to pinpoint the meaning of probabilities within multiverses. This is a pressing concern for Everettians because of the fundamental role played by probabilities and statistics in quantum mechanics. Simply put, to choose an interpretation of quantum mechanics unable to provide a straightforward account of the nature of probabilities is like to throw away the ladder one has just climbed from. This would make Everettian interpretations of quantum mechanics both more opaque and complex than their single-world counterparts and would provide a good reason to reject many-world interpretations as a whole.

In *One World versus Many*, Adrian Kent voices a variety of problems for Everettian interpretations which roughly center along those lines. Thus, he says: "My aim here is not to advocate a specific one-world version or variant of quantum theory, or to assess the current state of the art, but rather to compare and contrast one-world and many-world accounts of probability."[1] This is precisely the point we wish to discuss. We will not engage with the answers Kent provides to Greaves-Myrvold's and David Wallace's account of probabilities in terms of decision theory. Our whole concern is to refute the arguments Kent puts forward in support of the following claim : "No known version of the theory (unadorned by extra *ad hoc* postulates) can account for the appearance of probabilities and explain why the theory it was meant to replace, Copenhagen quantum theory, appears to be confirmed, or more generally why our evolutionary history appears to be Born-rule typical."[2] Kent thinks this problem is specific to many-world interpretations of quantum mechanics and can be avoided by their single-world counterparts through an alternative to the standard theory of probability. We think that Kent's substitution to probability theory faces important problems and that the worry he expresses about specific many-world interpretations of quantum mechanics are also problematic for single-world interpretations. We intend to show this through a series of rather mischievous thought-experiments.

---

[1] Kent, Adrian, « One World versus Many : the Inadequacy of Everettian Accounts of Evolution, Probability, and Scientific Confirmation », p. 3. Available online at http://arxiv.org/PS_cache/arxiv/pdf/0905/0905.0624v2.pdf.
[2] *One World versus Many*, p. 1.

# Act 1 - Guildenstern and Rosencrantz

Consider a simplified version of Kent's proposed cases.[3] Imagine a God—call him Tom Stoppard—ruling over a set of universes. Stoppard's universes are very simple. Their whole content is composed of an infinitely long blackboard, on which a string of cells are filled by single 0s or 1s, much like a Turing machine's tape. At some point, the numbers stop, however, as empty cells waiting to be filled extend into the distance. Besides this humongous blackboard, a duo of inhabitants—Guildenstern and Rosencrantz—quarrel about the meaning of the blackboard and a few other existential questions. The protocol through which Stoppard generates these numbers, elegantly named *Duplicate, Numerate and Destroy*, is straightforward. Every time He destroys some universe, two copies are made. In one copy, a 0 fills the first cell left empty on the Infinite Blackboard's tape. In the other universe, a 1 fills the same spot. This is the procedure Stoppard has followed religiously from the destruction of the first, primordial universe. The original facsimilation of this primordial universe in which the Infinite Blackboard was totally empty became the original step of *Duplicate, Numerate and Destroy*. This act marked the beginning of *The Age of Perplexity*.

Although adequately perplexed, Guildenstern and Rosencrantz are aware the entity known as Tom Stoppard destroys all universes every day and immediately creates a certain number of copies for each universe destroyed. How many? They don't know. They do know, however, that every time a universe is destroyed and then reduplicated, a new number—either 0 or 1—fills the next empty cell on each of the new universes' Infinite Blackboard. They have no access to the rest of the Lord's divine procedure. In every universe, copies of Guildenstern and Rosencrantz wonder what to think of this mess. Puzzled by their own predicament, they ask whether Guildenstern dies everytime Stoppard deletes the cosmos, whether Rosencrantz merely is the sum of his copies across the multiverse, or whether their immaterial souls fly from destroyed universes into freshly created one, animating their bodies and preserving their identities over time.

After some time, however, our philosophically bent inhabitants agree on turning to more pragmatic questions and try to infer Stoppard's real protocol. Given their lack of information, Guildenstern and Rosencrantz assume He creates a number $n$ of universes with a 0, and a number $m$ of universes with a 1 every day. They ask: "What is the proportion $p$ of universes created with a 0 ($p = n/n+m$) and what is the proportion $q$ of universes created with a 1, ($q= 1-p$)?" How could they answer this question? Assume they turn to frequentism. In such a scenario, the Guildensterns and Rosencrantzs of Stoppard's multiverse will estimate the proportion of universes created with a 0 as the proportion of 0s inscribed so far on their own blackboard. They will revise this estimation every day, whenever a new number appears on the Turing-like tape.

When Guildenstern and Rosencrantz ask what proportion of universes have a certain property, this defines a probability measure in the traditional mathematical sense of the term. This probability measure satisfies Kolmogorov's axioms. First, the proportion of the totality of universes is one, no set of universes has a negative proportion, and the total proportion of two disjoint sets of universes is equal to the sum of the proportion of both sets. Thus, all mathematical proofs relying on statistics and probabilities apply to this toy model. We are justified to say this probability space will reflect the experience of Guildenstern and Rosencrantz for reasons of symmetry: all universes are created equal, they differ only by the addition of either a 0 or a 1 on the blackboard. These numbers have no further metaphysical or physical significance. However, some people might think that a mathematical probability does not necessarily translate into a more intuitive, subjective sense of probability. This is not a problem for us. More will be said about this later.

---

[3] In Kent's original article, simulated beings are pressing a red button to create new universes. We modified this in our presentation because then some branches might simply never happen: the simulated beings might decide never to press to red button given certain circumstances. And then Tom Stoppard would not be in the complete control of the situation.

In such a situation, frequentism is a completely reasonable strategy. In the absence of further information about Tom Stoppard's motives, habits or beliefs, Guildenstern and Rosencrantz cannot justify probabilistic *a prioris* about Stoppard's protocol. The Lord is, after all, a transcendent being. Because His nature cannot be inferred from the inhabitants' surroundings, all talk of probabilistic weights not entirely derived from observed frequencies is irrelevant to Guildenstern and Rosencrantz's concerns. Without a reliable form of Divine Revelation coming down from the sky, all the relevant information accessible to our inhabitants stands on the Infinite Blackboard. Nothing else matters. We say this in reply to Kent's discussion of branch weight. In some of Kent's toy universes, various weights are attached to branches. Transposed in our theatrical scenario, Kent's weights should work like this. Imagine Guildenstern and Rosencrantz find a strange bottle in which they find the following message: *n=35, m=65*. Without any *a priori* about the message's truth, Guildenstern and Rosencrantz cannot assume it constitutes an important piece of information. Either the message gives the real values of *m* and *n*, or it does not. If it does, its accuracy of the value of *p* and *q* can be determined by any typical Guildenstern simply looking at the Infinite Blackboard. If it doesn't, again, looking at the Infinite Blackboard does the trick. The message has no informational value. Given the circumstances, frequentism is the best strategy.

Nonetheless, some problems remain unavoidable. In a very small minority of universes, an atypical sequence stands on the Infinite Blackboard, that is, a sequence unrepresentative of the rules followed by Tom Stoppard. In those universes, our protagonists find themselves in a predicament very similar to the situation described in the real Tom Stoppard's play *Rosencrantz and Guildenstern Are Dead*, in which consecutive coin-tosses yield 92 heads in a row. This is a highly atypical sequence, one which—although logically possible—is unrepresentative of the probabilistic rules of coin-tossing. If the real Tom Stoppard's Rosencrantz and Guildenstern were to generalize their obtained statistics into a probabilistic theory about coin-tossing, they would be widely off the mark. To adopt frequentism is to assume that the sequence initially given is a typical sequence and that, as the number of trials increases, the sequence obtained will be more and more representative of the real probability underlying trial results.

Problematically, frequentism never avoids the logical possibility of very long, persistent atypical sequences. Given a large enough number of universes within a multiverse, there will be some atypical universes in which data leads to erroneous conclusions. In other words, by adopting frequentism, some duos of Guildensterns and Rosencrantzs will be led astray in their quest for the Divine Protocol. Kent rightly argues there is no way to know if one's own universe is typical or atypical. However, he also thinks this is a crippling argument against Everettian quantum mechanics. We think, on the other hand, that this is rather a fundamental limitation of probabilities and statistics. Such worries are in no way more crippling for many-worlds interpretations of quantum mechanics than for their single-world counterparts. Or indeed, even for the simplest coin-toss experiment. This problem merely is the problem of induction, a problem which, although worrisome in theory, is perpetually overcome in practice.

# Act 2 - Guildenstern and Rosencrantz in Space

Imagine a different scenario, taking place in a single, gigantic universe. This universe contains $2^{1\,000\,000\,000}$ planets, all inhabited by copies of our Shakespearian duo[4]. Each planet is endowed with a colossal but finite blackboard with a tape-like inscriptions of 0s and 1s. Each day, a new cell is filled on the tape while all previous inscriptions remain the same. During the first day of this single universe's existence, exactly half the planets obtain a 0 and the remainder obtains a 1. The second day, exactly a quarter of the planets harbour each possible two-bit sequence and so on. Guildenstern and Rosencrantz witness this process until, after a billion days, new digits stop appearing on each planet's blackboard. Alarmed and puzzled by such an unexpected event—which they agree on calling the *Day of Ultimate Computation*—Guildenstern and Rosencrantz are persuaded their own salvation depends on producing Stoppard's protocol in front of the Great Ruler himself.

Imagine that Guildenstern, terrified by the prospects of failure, laments in despair: "Useless! Why inquire about any Supreme Being's whim when this Divine Withholder relishes at the sight of his own creatures' frenzied efforts to conform to some undisclosed law? Malice is this sole Creator's motive, and I would wage against there being any Divine Protocol at all. No law explains or gives rise to these cursed numbers! Only blind chance and pure arbitrariness! That, and the ecstasy enjoyed by He who cherishes the sight of fools staking their lives on the existence of such fictions!" And so, convinced the emergence of digits on the blackboard is a pure random event, simply designed to cause them additional torment before their final demise, Guildenstern buries his face in his hands and abandons all hope.

A more resolute Rosencrantz could easily reply : "You call Divine Law a fiction because you perceive malice in our Lord's mysterious silence. But have you thought of the sheer absurdity of your words? An event foreign to cosmic order and devoid of origin! You would like to call such a bastard child of chance "Random". But what exact thought corresponds to this lame pseudonym? Can you imagine a single process caused by this snot-faced brat? Is causation even within reach of his greasy hands? No, my friend, when you claim no protocol explains what our eyes have seen since this planet came into being, I can only think the despair in your heart dictates the capitulation of your mind."

As Rosencrantz steps down from his soapbox, Guildenstern would probably realize this alternative really is enigmatic. How can we wrap our heads around true randomness, understood as a real feature of the world and not merely a by-product of ignorance? More importantly, what could one possibly achieve with such a presupposition? From a pragmatic point of view, to assume data stems from a completely and pervasively random source is simply to assume there is nothing to be learned or predicted about a given phenomenon. Such a principle is sterile. Imagine Guildenstern realizes the error of his ways and agrees with Rosencrantz to figure out Stoppard's protocol. To what tool do they turn? Standard probability theory and frequentist statistics once again. Much like in our multiverse case, a few planets harbour atypical sequences, while the overwhelming majority of planets have typical sequences of 0s and 1s. As before, the inhabitants of typical planets possess the best understanding of what is going on elsewhere and the greatest ability to understand their own future. They come very near to Tom Stoppard's real protocol. Also, in most planets where inhabitants observe atypical results, this atypicality is short-lived and dissolves over time.

This case is rigorously analogous to the previous one, even if it takes place within a single universe. This shows that Kent's problem, which focuses on the impossibility of knowing with certainty whether one's data is representative of the laws of the universe, is not a problem specific to Everettian interpretations of quantum mechanics, but applies just as well to single universe theories. In single universes just as in multiverses, no one can be absolutely certain one's results are typical. But what else is there to do, except assume typicality and wait for further results? Give up the idea that the past events should somehow inform us about the future? This seems to be the only alternative, one which is unacceptable in theory, and inefficient in practice. Let's not follow Guildenstern's footsteps here and bury our head in our hands in desperation.[5]

---

[4]We challenge any reader who thinks such a number of planets within in a single universe is impossible to provide a *deductive* proof for this claim.

## *Exeunt*—Biased Coin Reinterpretation and Deductive Falsifiability

The lack of a mechanism deductively justifying theory refutation on the basis of observation is one of the most persistent problems addressed in Kent's article. While traditional statistical tools allow inhabitants of typical planets and universes to tentatively "refute" and "prove" theories, they are useless to the very small minority living in weird universes or planets. These standard tools will systematically lead them to "prove" and "refute" the wrong theories. And since there is no way of knowing whether we live in an atypical branch of Stoppard's multiverse, Kent thinks this impossibility invalidates the use of these traditional tools in the context of many-worlds theories of quantum mechanics. In other words, multiverse inhabitants would not be justified to infer probabilities out of statistics. If true, this would be a major problem for Everettian interpretations of quantum mechanics. Kent also thinks this problem can be sidestepped in single-world interpretations and argues for the superiority of such interpretations on account of this difficulty. He defends this thesis by proposing a model which—he thinks—would make such purely deductive refutation possible in single universes. We will argue he is mistaken on both points. Kent says:

> The theory PT [a given probabilistic theory] is thus not *logically refuted* by the outcome S [an extremely improbable event according to the theory]. In practice we would reject it—but, without a fundamentally satisfactory account of probability, it is hard to give a completely satisfactory justification for doing so. In our alternative account, however, no such problem arises. [...] On this view of scientific accounts of apparently random data, that's the best one can hope for: generically, no single clearly optimal theory will emerge. However, we can hypothesize that theories of roughly this length are essentially best possible—i.e. that the string cannot be compressed to significantly shorter than $H(p)N$ bits—and this hypothesis *is testable and falsifiable*.[6]

Kent claims to provide an account of probability which yields not mere inductive justification for rejecting a given probabilistic theory on account of conflictual observations, but a full-blown deductive one. This is what he means by a *logically refutable theory*. The endeavour is far from trivial. If Kent succeeds, not only would he solve an age-old philosophical problem, but he would also provide a theoretical reason to prefer single-world interpretations of quantum mechanics to their multi-world counterparts. We think, however, that Kent's account fails to meet these standards.

Let us follow Kent for a moment. Imagine a source producing strings of 0s and 1s, strings that look like the results of fair coin tosses. This process is apparently random, although possibly pseudo-random. We observe a large string of 0s and 1s. Using traditional means, even if we observe 1 % of 0s and 99% of 1s, we can never deductively refute the hypothesis that the coin is fair. This will remain true independently of string length. Standard probability theory allows for the possibility that a fair coin could produce these results. Thus, as Kent says, we might have practical justification for rejecting the hypothesis that the source behaves like a fair coin, but this justification falls short of the absolute certainty that would provide deductive justification.

---

[5]The proportion of planets with a certain property defines a probability space just like in our previous example. The same question still applies if we ask whether this mathematical space translates into an intuitive notion of physical probability.
[6]*One World versus Many*, p. 5-6.

Kent proposes a rather complex alternative. He builds an account of theory refutation on the basis of Shannon entropy. However, Shannon entropy is not what warrants his claim to deductive falsifiability. Therefore, we refer any reader interested in this point to Kent's own article.[7] We will stick to the essential feature of Kent's account of deductive falsifiability. Kent starts by considering a very large number of hypotheses (almost every possible one, rejecting a few for criterions of simplicity and elegance). For any single theory according to Kent's account, some sequences of events lie within the predictive scope of a theory. When an observed event does not figure within that predictive scope, the theory is logically refuted. The crucial step of Kent's alternative comes when he excludes some very improbable events, but events which are nonetheless still possible according to standard probability theory:

> Again, [Kent's] theories reproduce deterministically predictions that the standard probabilistic theory says hold with probability very close (but not equal) to one. They exclude some very low probability events which would, if realised, in practice persuade almost everyone that the probabilistic theory was wrong, even though their occurrence is logically consistent with the theory.[8]

This way of treating very improbable but not altogether impossible events might indeed be an advantage proper to single-world theories of quantum mechanics. Since according to multi-world theories, every branch is actual, to disregard weird branches is not merely to make abstraction of some rare possibility, but to actually fail to take into account something real. However, even if making abstraction of very weird, improbable events seems to be the prerogative of the single-world theorist, we think such an abstraction is unwarranted and leads to important errors.

Where does this desire for deductive falsifiability come from? From Popper's philosophy of science. If science works from induction, there is no deductive basis for accepting theories because no inductive argument ever attains deductive certainty. Where does scientific validity come from if theory confirmation has no logical basis? Popper's answer to this problem was that science does not prove theories through induction, but rather falsifies them through deduction. A theory's scientificity resides in its falsifiability. Accordingly, what made Newtonian mechanics scientific is the fact that this theory could be refuted. Indeed, it has. Relativity theory now stands in its place. This case provides an illustration of Popper's understanding of scientific progress: a single experiment can falsify a deterministic physical theory:

> The acceptance by science of a law or of a theory is tentative only; which is to say that all laws and theories are conjectures, or tentative hypotheses (a position which I have sometimes called 'hypotheticism') [...] So long as a theory stands up to the severest tests we can design, it is accepted; if it does not, it is rejected. But it is never inferred, in any sense, from the empirical evidence. There is neither a psychological nor a logical induction. Only the falsity of the theory can be inferred from empirical evidence and this inference is a purely deductive one.[9]

But not so fast. We believe this conception of scientific progress is simplistic. Although it looks good to the logician, no empirical scientist ever employs it. Suppose a coyote drops an anvil off a cliff (perhaps aiming for some fast-moving bird). Suppose the anvil immediately flies up only to come down smashing on the poor, hungry and puzzled animal. Is universal gravity thereby falsified? No. This would be reckless science. No one can assume that experimental settings have ruled out all external factors. There could be a huge, powerful magnet counterbalancing gravitational force somewhere in the sky. Two different problems can be found with such an account. On the one hand, Popper's theory only applies to toy models of deterministic sciences in which all factors are under precise control. Remember that deductive entailment supposes *logical certainty*. To claim a single experiment can invalidate a whole theory is to claim that it is logically impossible that something was wrong with one's experimental protocol or observations. No scientist can ever claim his protocol to be that smooth —never with this degree of certainty. Popper did recognize this fact, and merely proposed deductive falsifiability as a criterion of what is empirical or not. In other words, for any possible empirical theory, there should in principle be some observation which would deductively invalidate the theory. In practice, however, things are more complex. At the time when this article was being written, the scientific community was shocked by the apparent discovery that neutrinos can travel faster than the speed of light. This discovery could refute Einstein's relativity theory. However, scientists are still struggling to find out what really happened. This shows that, in practice, a single or even a few observations do not lead naturally to deductive refutation. How many experiments do we need? With this question, we come back face to face with our old friend induction. Falsifiability—in practice and within deterministic sciences—does not work deductively but inductively.

---

[7]*One World versus Many*, p. 4-6.
[8]*One World versus Many*, p. 6.
[9]Karl Popper, "The Problem of Induction," *in Popper Selections*, ed. David Miller, Princeton University Press, 1985, p. 102.

A second problem exists with Popper's account. There are sciences which are not overtly deterministic, either because of pervasive pseudo-randomness or true randomness. Pharmaceutical science in particular deals with evidence that is irremediably statistical in nature. No single event will ever count as deductive refutation in pharmaceutical science because its predictions make room for weird, atypical events. Drugs do not always have the same effects on everyone. The same is true for quantum mechanics, and perhaps even more so because there are good reasons to think that pharmaceutical science is based on merely apparent randomness, while the randomness of quantum mechanics is of a very different and elusive nature.

Therefore, according to this criterion of what is scientific or not—deductive falsifiability— pharmaceutical science would not qualify as an empirical science because no single observation could deductively refute our theories about the effectiveness of a given drug. Not even in principle. The same applies to quantum mechanics. The tools of deductive logic which are the hallmark of theoretical computer science and mathematics are certainly useful in empirical science. However, they do not suffice in establishing the scientific validity of some theory. Falsifiability works inductively in practice.

## Act 3 - Guildenstern and Rosencrantz's Great Falsification

We propose a new thought experiment in order to engage Kent's theory on his own grounds. Imagine copies of Guildenstern and Rosencrantz trapped in space, once again. Our sympathetic duo lives, as before, in a single universe with an enormous amount of planets. In an act of Divine Providence, Tom endows all of them with rigorously identical quantum automatons.[10] These automatons are identical down to the atomic level. Suppose they interpret this quantum analytical engine's surnatural appearance as a sign of some divine command—*Thou shall determine the divine automaton's fairness —where fairness is defined as a 50% chance of obtaining 0 and a 50% chance of obtaining a 1*. Furthermore, suppose our rather paranoid inhabitants believe a wrong answer will trigger the *All-Branching Apocalypse*—a cataclysmic scenario in which the universe would divide in four branches. In each of these branches, copies of Guildenstern and Rosencrantz would be hunted down by one of the four horsemen of the Apocalypse. In fact, no such thing will happen. The universe will not and cannot branch, and the horsemen of the Apocalypse are mere fairy tales. However, because of this belief, an infallible method would cause Guildenstern and Rosencrantz great relief.

---

[10]For the few readers not familiar with forbidden engineering and quantum alchemy, a quantum automaton can be thought of as a steam-powered device with a copper lever, and an extremely large Turing tape, as in our previous single universe example. Each time the copper lever is pulled, either a 0 or a 1 will mysteriously appear on the tape, on the rightmost empty cell. It was invented by Lady Quanta Lovelace in an alternate universe.

Suppose our Celestial Dramatist ensures the statistics obtained on all tapes are rigorously identical to the statistics in our previous single universe thought-experiment. After pulling the automatons' copper lever a few million times, a great majority of Guildensterns and Rosencrantzs will come to the conclusion that the quantum automaton is fair. However, given the formidable amount of planets where trials are conducted, weird results are bound to happen. As the inhabitants of this universe are aware of the number of planets and the number of people like them living on these planets, imagine the following dialogue occurs on a planet where the ratio of 0s to 1s is 9:1.

***

*Guildenstern*: "Alas, my poor Rosencrantz! How ardent is my wish for indubitable truth! How miserably do our theories squirm in the face of absolute certainty! Unavoidably, someone, somewhere across this universe is bound to fail through our very method. This poor soul will fail without stumble or miscalculation, only through the brute imperfections of chance!"

*Rosencrantz*: "Take hope, Guildenstern! I have found an arcane procedure invented by some quantum wizard, a procedure yielding infallible theory-refutation. Because only two hypotheses compete for our approbation, deductive falsification here means indubitable truth. Let us discover whether the Sacred Automaton is reliable or another of His wicked tricks! Although the formula is quite complex, the main steps are straightforward: 1) assume typicality and 2) exclude extremely improbable events as logically impossible."

*Guildenstern:* "My dear Rosencrantz! Redemption is ours if this quantum charm is as potent as its brewer claims. You and I will never run eternal circles in front of horse-riding Death or saddled Sickness. How does this quantum wizard prove the logical impossibility of the extremely improbable?"

*Rosencrantz*: "Nowhere. Assuming does the trick. 'Tis sufficient."

*Guildenstern*: "Surely, this will not do. Magic should be more elaborate."

*Rosencrantz*: "Guildenstern, redemption is handed to you on a plate. Can your stomach really be bothered by something so desirable? Disgust is most inappropriate in such circumstances."

*Guildenstern*: "Think, my friend. This arcane procedure only grants us redemption through sound assumptions. How can this wizard turn the merely improbable into the logically impossible? Such incantations are unheard of. We are warranted to assume we live on typical ground and lead typical lives. We are justified in thinking the extremely improbable will not happen here. Not to us. But never with absolute certainty. Why assume logic allows special treatment in this part of the universe? This land, our lives, these things mean nothing to logic's iron hand."

*Rosencrantz*: "Guildenstern, I hear your words, but I fear they are parting words. I will use the quantum wizard's formula. I have no use for anything less than absolute certainty. Do according to your wish, I will stand by my resolution."

***

Imagine that, once Guildenstern and Rosencrantz have agreed to disagree, Tom sends a steam-powered space zeppelin (expertly piloted by an archangel) which allows them to visit every other planet (given astronomical time) and verify the numbers given by other automatons. In doing so, they obtain proof their own planet was atypical and that the quantum automaton was, in fact, fair. They also observe that, on typical planets, the merely inductive method followed by Guildenstern led their copies to answer correctly. This comes as a shock to Rosencrantz, who was convinced being wrong was logically impossible. To be sure, Guildenstern is also surprised to find out he lived in a profoundly atypical world, but his method allowed such a possibility. Since the automatons are identical down to the atomic level, Rosencrantz is then forced to either reject the validity of his own procedure—which was also Kent's—or deny that atomic level symmetry is a sufficient reason to believe the automatons were strictly equivalent.

This is an important problem for Kent's account of theory-refutation. Either Kent needs to say the theory selected by his theory-refutation procedure is only applicable to the particular source which yielded his original data, or these results can be extended to relevantly similar sources. In the former case, Kent's proposal would not amount to anything remotely scientific in nature. If one refuses to derive predictions about relevantly similar sources, objects or events, then one refuses to subject one's theory both to the main test of scientific validity and the main practical use science has in general. This would be tantamount to saying, in the context of quantum mechanics, that a given theory does not say anything about equivalent particles, but only about *this* particle. This is obviously problematic. To refuse to make predictions about similar sources, objects or events is to refuse to engage in science altogether.

If Kent is willing to use his deductively obtained theory to provide predictions about the behaviour of relevantly similar sources, objects or events, then he will have to recognize that sources may be equivalent without giving remotely similar data. Similar particles can yield radically different results given enough trials but this is no foolproof reason to think our generalizations do not apply to all these particles. To recognize that fact is to recognize that no process of deductive refutation can be extrapolated from a single source and then applied to further observations. This is clearly in conflict with Kent's proposal. He says: "Conversely, the theory logically (not merely with high probability) implies that we will see no consistent regularities in our experimental data that would, on the usual account, be highly improbable."[11] This, we argue, is incompatible with the basic scientific task which consists in providing predictions about the behaviour of relevantly similar objects. Even if, after a very large number of trials on a given type of particle, we have never observed a single particle which behaved as atypically as one currently under observation, we cannot be justified in thinking one could deductively refute the hypothesis that this particle is like all previous ones. With enough trials, even equivalent sources are bound to exhibit great variation. This is a burden inherent to scientific explanation.

One final point: Kent's proposed procedure is non-standard in that it starts with almost every possible hypothesis about the structure of data (almost, because some are excluded for reasons of simplicity and elegance). Accordingly, even very large amounts of data will never succeed in narrowing down this melting-pot of hypotheses into a single, usable theory. Think about poor Guildenstern and Rosencrantz, which are concerned about their own future and the future of their universe. How can they predict anything about their own future given millions of theories? Science is a practical endeavour which needs a single or a few models of the world in order to make predictions. Having a few millions theories is practically similar to having no hypothesis at all. If Kent's account of theory-refutation does not select in the end one or two theories which are considered the right ones, our human protagonist will have to do it anyway for practical reasons.

---

[11] *One World versus Many*, p. 6.

# Epilogue

Progress in philosophy is often not about finding final answers but about converging toward the real nature of some enduring problem. Philosophical problems about quantum mechanics make no exception. In our theatrical universes final answers are Tom Stoppard's privilege. They are inaccessible to those who inhabit His tales. Tom's creatures can only approximate. They can never know the Celestial Playwright's true intentions. This process of approximation, however, is remarkably reliable. In reply to Kent, we think we have shown how atypicality is not a problem specific to many-world interpretations of quantum mechanics. This problem merely is the problem of induction, an enduring problem for the foundations of science which cannot be superseded in theory but can be overlooked in practice. Science cannot be reduced to logic and deduction will never replace induction. If science was reducible to logic, no prediction could ever be made. Statistics and probability are unavoidable tools in science.

**Acknowledgments** We thank Gilles Brassard and Alain Tapp for their precious help, discussion and review of earlier drafts of this paper.